\documentclass[twoside]{article}

\usepackage[accepted]{aistats2024}
\usepackage{booktabs}       
\usepackage{graphicx} 
\usepackage{caption} 
\usepackage{rotating}
\usepackage{nicefrac}       
\usepackage{microtype}
\usepackage{bm}
\usepackage{amsmath}
\usepackage{amsthm}
\theoremstyle{definition}

\usepackage{cleveref}
\usepackage{natbib}
\usepackage{xcolor}

\setcitestyle{numbers,sort,round}
\setcitestyle{authoryear,open={(},close={)}, aysep={,}} 
\usepackage{subcaption}

\usepackage{amssymb}
\usepackage{pifont}
\newcommand{\cmark}{\ding{51}}%
\newcommand{\xmark}{\ding{55}}%
\newcommand{\modelabbrev}{DISEE}
\newcommand{\modelname}{{D}ynamic {I}mpact {S}ingle-{E}vent {E}mbedding Model}
\begin{document}

%

\runningtitle{Time to Cite: Modeling Citation Networks using the \textsc{\modelabbrev} Model}

%
\runningauthor{Nakis, \c{C}elikkanat, Boucherie, Lehmann, Mørup}

\twocolumn[

\aistatstitle{Time to Cite: Modeling Citation Networks using the Dynamic Impact Single-Event Embedding Model}

\aistatsauthor{ 
Nikolaos Nakis \And Abdulkadir \c{C}elikkanat \And  Louis Boucherie }
\aistatsaddress{Technical University \\of Denmark \And Technical University \\of Denmark \And Technical University \\of Denmark}

\aistatsauthor{ 
Sune Lehmann \And Morten Mørup }
\aistatsaddress{Technical University \\of Denmark \And Technical University \\of Denmark}]

\begin{abstract}
  Understanding the structure and dynamics of scientific research, i.e., the science of science (SciSci), has become an important area of research in order to address imminent questions including how scholars interact to advance science, how disciplines are related and evolve, and how research impact can be quantified and predicted. Central to the study of SciSci has been the analysis of citation networks. Here, two prominent modeling methodologies have been employed: one is to assess the citation impact dynamics of papers using parametric distributions, and the other is to embed the citation networks in a latent space optimal for characterizing the static relations between papers in terms of their citations. Interestingly, citation networks are a prominent example of single-event dynamic networks, i.e., networks for which each dyad only has a single event (i.e., the point in time of citation). We presently propose a novel likelihood function for the characterization of such single-event networks. Using this likelihood, we propose the \textsc{D}ynamic \textsc{I}mpact \textsc{S}ingle-\textsc{E}vent \textsc{E}mbedding model (\textsc{\modelabbrev}). The \textsc{\modelabbrev} model characterizes the scientific interactions in terms of a latent distance model in which random effects account for citation heterogeneity while the time-varying impact is characterized using existing parametric representations for assessment of dynamic impact. We highlight the proposed approach on several real citation networks finding that the \textsc{\modelabbrev} well reconciles static latent distance network embedding approaches with classical dynamic impact assessments.
\end{abstract}

\section{Introduction}\label{sec:introduction}

The abundance of scientific data has established the science of science (SciSci) as a vital tool for understanding scientific research, as well as predicting future outcomes, research directions, and the overall evolution of science \citep{scisci_review}. More specifically, SciSci studies the methods of science itself, searching for answers to important questions such as how scholars interact to advance science, how different disciplinary boundaries are removed, and how research impact can be quantified and predicted. SciSci is an interdisciplinary field with various prominent research directions including but not limited to, scientific novelty and innovation quantification \citep{nov1,nov2,nov3,nov4,nov5}, analysis of career success dynamics of scholars \citep{car1,car2,car3,car4,car5,car6,car7}, characterization of scientific collaborations \citep{col1,col2,col3,col4,col5} as well as citation and research impact dynamics \citep{imp1,imp2,imp3,imp4,imp5,imp6,imp7}.

A major focus has been given to the understanding of SciSci through the lens of complex network analysis, studying the structural properties and dynamics, of the naturally occurring graph data describing SciSci. These include collaboration networks describing how scholars cooperate to advance various scientific fields. In particular, pioneering works \citep{sci_col1,sci_col2,sci_col3} have analyzed various network statistics such as degree distribution, clustering coefficient, and average shortest paths. Furthermore, citation networks define an additional prominent case where graph structure data describe SciSci. Citation networks, essentially describe the directed relationships of papers (nodes) with an edge occurring between a dyad if paper $A$ cites paper $B$, e.g. A$\rightarrow$B. Studies focusing on citation networks have shown power-law and exponential family degree distributions \citep{cit_net1}, sub-field community structures \citep{cit_net2}, and tree-like backbone topologies \citep{cit_net3}. Lastly, bipartite network structures can emerge by defining networks describing author-paper relationships, including indirect author connections through a collaboration paper or through their citing patterns \citep{auth_pap1,auth_pap2,auth_pap3}.

In this paper, we focus on citation networks that allow for paper impact characterization. Notably, such networks are directed and dynamic ideally having an upper triangular adjacency matrix when nodes are sorted by time due to the time-causal structure of citations (i.e., new papers can only cite past papers, where rows denote cited papers and columns citing papers). 

Initial works for paper impact quantification utilized classical machine learning methods on various scholarly features, as well as paper textual information. Methods used to estimate future citations included linear/logistic regression, k-nearest neighbors, support vector machines, random forests, and many more \citep{feat1,feat2,feat3,feat4,feat5,feat6}. These studies focused primarily on carefully designing and including proper features (H-index, impact factor (IF), etc.) to be used for the impact prediction task. 
While these methods attracted lots of attention, they have a major limitation where papers with very similar features define much different citation distributions and attention patterns that are not characterized.

Later works tried to define impact on the paper level by treating the accumulation of citations through time as a time series. In the original work of \citep{ts1}, Redner proposed a log-normal distribution to fit the cumulative citation distribution for papers published during a $110$ year period in Physical Review. This was followed by \citep{ts2} using a shifted power-law distribution on the same networks. Furthermore, another widely used distribution modeling citation dynamics is the Tsallis distribution proposed in \citep{ts3}. The log-normal and Tsallis distribution share a lot of similarities but in literature, the log-normal is preferred due to its simplicity. Later works combined the important characteristic of preferential attachment with the log-normal distribution \citep{ts4,ts5,ts6}, as well as, the Poisson process \citep{ts7,ts8}.


Various prominent Graph Representation Learning \textsc{(GRL)} methods have also been applied to citation graphs \citep{cora,hepth} as they have been very popular network choices for assessing downstream task performance, such as link prediction and node classification \citep{deepwalk-perozzi14,node2vec-kdd16,netmf-wsdm18}. Recently, Graph Neural Networks \textsc{(GNN)}s have also been used including GraphSAGE \citep{graphsage_hamilton}, the Adaptive Channel Mixing \textsc{GNN} \citep{luan2021heterophily} and Convolutional Graph Neural Networks \citep{kipf2017semisupervised}. Despite these works defining strong models, powerful link predictors, and node classifiers they do not explicitly account for impact characterization, nor for the dynamic way that paper citations appear.

\begin{figure*}[!t]
\centering
\includegraphics[width=\textwidth]{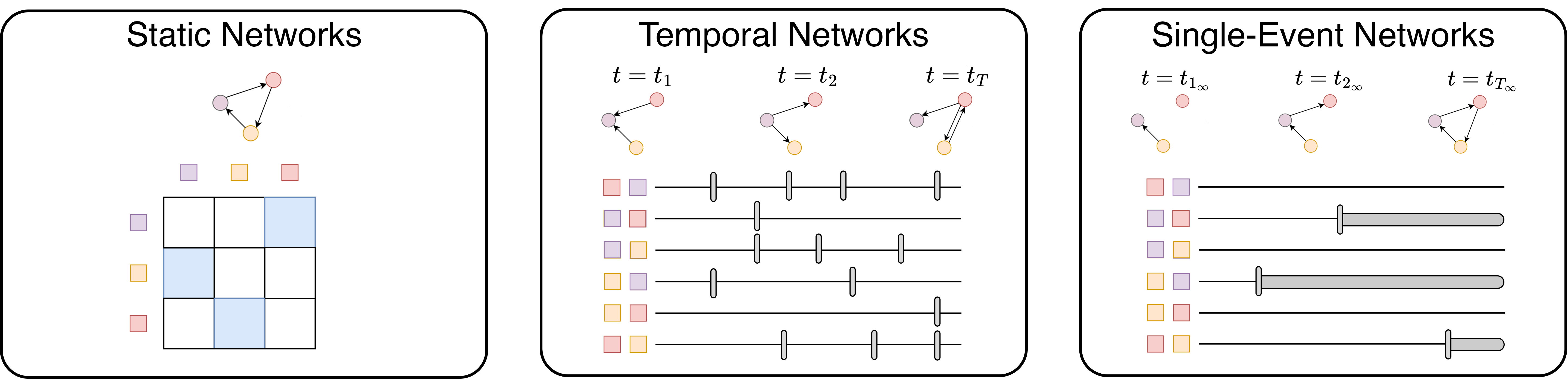}
\caption{Examples of three different types of networks based on their temporal structure. Round points represent network nodes, square points make up the corresponding colored node dyads, arrows represent directed relationships between two nodes, vertical lines represent events, and black lines are the timelines while grey bold lines show that a link (event) appeared once and cannot be observed again. \textit{Left panel:} Static networks where links occur once and there is no temporal information available. \textit{Middle panel:} Temporal networks where links are events in time and can be observed multiple times along the timeline. \textit{Right panel:} Single-event networks (SENs) where links appear in a temporal manner but can occur only once for each dyad, defining edges as single events. }\label{fig:sen}
\end{figure*}

Notably, citation networks are dynamic. Whereas dynamic modeling approaches can uncover structures obscured when aggregating networks across time to form static networks, the dynamic modeling approaches are in general based on the assumption that multiple links occur between the dyads in time. Importantly, for continuous-time modeling, this has typically been accounted for using Poisson Process likelihoods \citep{hawkes_1,fan2021continuous, celikkanat2022piecewise,celikkanat2023continuoustime} including likelihoods accounting for burstiness and reciprocating behaviors by use of the Hawkes process \citep{hawkes_1,hawkes_2,hawkes_3,htne,fan2021continuous}. 
To account for the high degree of complex interactions in time, advanced dynamic latent representations have further been proposed considering both discrete-time \citep{ishiguro2010dynamic,herlau2013modeling,heaukulani2013dynamic, deepwalk-perozzi14, node2vec-kdd16, durante2014bayesian, durante2016locally,kim2018review, ldm_1,ldm_2} and continuous-time dynamics \citep{hawkes_1,hawkes_2,hawkes_3,fan2021continuous,celikkanat2022piecewise,celikkanat2023continuoustime}, including \textsc{GNN}s with time-evolving latent representations \citep{dyrep, gnn_1}. For surveys of such dynamic modeling approaches see also \citep{survey1,survey2}. 

Importantly, citation networks are a class of dynamic networks characterized by a single event occurring between dyads. We presently denote such network a Single-Event Network (\textsc{SEN}). I.e., links occur only once at the time of the paper publication. However, neither of the existing dynamic network modeling approaches explicitly account for \textsc{SEN}s. Whereas continuous-time modeling approaches are designed for multiple events, thereby easily over-parameterizing such highly sparse networks, static networks can easily be applied to such networks by disregarding the temporal structure but thereby potentially miss important structural information given by the event time. Despite these limitations, to the best of our knowledge, existing dynamic network modeling approaches do not explicitly account for single-event occurrences. In Figure \ref{fig:sen}, we provide an example of three cases of networks that define static, traditional event-based dynamic networks, as well as \textsc{SEN}s. We here observe how static networks are completely blind to the temporal information that single-event networks capture while it is also evident that they differ from traditional event-based temporal networks where each dyad can have multiple events across time.

When modeling \textsc{SEN}s, the single event occurrence makes the networks highly sparse. To account for the high degree of sparsity of SENs we use as a starting point the static Latent Distance Modeling (LDM) approaches proposed in \citep{exp1} in which static networks are embedded in a low dimensional space and the relative distance between the nodes used to parameterize the probability of observing links between the nodes. Importantly, these modeling approaches have been found to provide easily interpretable low-dimensional ($D=2$ and $D=3$) network representations with favorable representation learning performance in tasks including link prediction and node classification  \citep{nakis22hbdm,HMLDM}. The LDM has been generalized to distances beyond Euclidean, including squared Euclidean distances and hyperbolic embeddings \citep{nickel2017poincare,nickel2018learning} as well as to account for degree heterogeneity through the use of node-specific biases (denoted random effects) \citep{doi:10.1198/016214504000001015,KRIVITSKY2009204,nakis22hbdm} which we presently refer to as the mass of a paper.
Notably, we define paper masses based on their citation dynamics through time, regulated by their distance in a latent space used to embed the structure of the citation network. Specifically, to account for single-event network dynamics, we endow the cited papers (receiving nodes) with a temporal profile in which a parametric function as used for traditional paper impact assessment \citep{ts1} is employed to regulate the nodes' citation activity in time forming the \textit{\modelname} (\textsc{\modelabbrev}). In particular, our contributions are:

\textbf{1) We derive the single-event Poisson Process (SE-PP).} As paper citation networks only include a single event we augment the Poisson Process likelihood to have support only for single events forming the single event Poisson Process.

\textbf{2) We propose the \textsc{\modelabbrev} model based on the SE-PP for \textsc{SEN}s.} We characterize the rate of interaction within a latent distance model augmented such that citations are generated relative to the degree to which a paper cites and a paper is being cited at a given time point interpreted as masses of the citing and cited papers in which the mass of the cited paper is dynamically evolving.
 
 \textbf{3) We demonstrate how \textsc{\modelabbrev} reconciles conventional impact modeling with latent distance embedding procedures.} We demonstrate how \textsc{\modelabbrev} enables accurate dynamic characterization of citation impact similar to conventional paper impact modeling procedures while at the same time providing low-dimensional embeddings accounting for the structure of citation networks. We highlight this reconciliation on three real networks covering three distinct fields of science.

The paper is organized as follows. In Section \ref{sec:SE-PP}, we present the single-event Poisson Process (SE-PP) for the modeling of single-event networks (\textsc{SEN}s). In Section \ref{sec:DIC} we discuss procedures for characterizing dynamic impact, and in section \ref{sec:DISEE}, we demonstrate how existing embedding procedures can be reconciled with dynamic impact modeling using the SE-PP by the proposed \textsc{D}ynamic \textsc{I}mpact \textsc{S}ingle-\textsc{E}vent \textsc{E}mbedding Model (\textsc{\modelabbrev}). In Section \ref{sec:results}, we present our results on the three distinct citation networks contrasting the performance to the corresponding conventional impact dynamic modeling, as well as, the powerful static \textsc{LDM} \citep{nakis22hbdm}. Section \ref{sec:conclusion} concludes our results.


\section{The Single-Event Poisson Process}\label{sec:SE-PP}

\begin{figure*}[!t]
\centering
\includegraphics[width=0.8\textwidth]{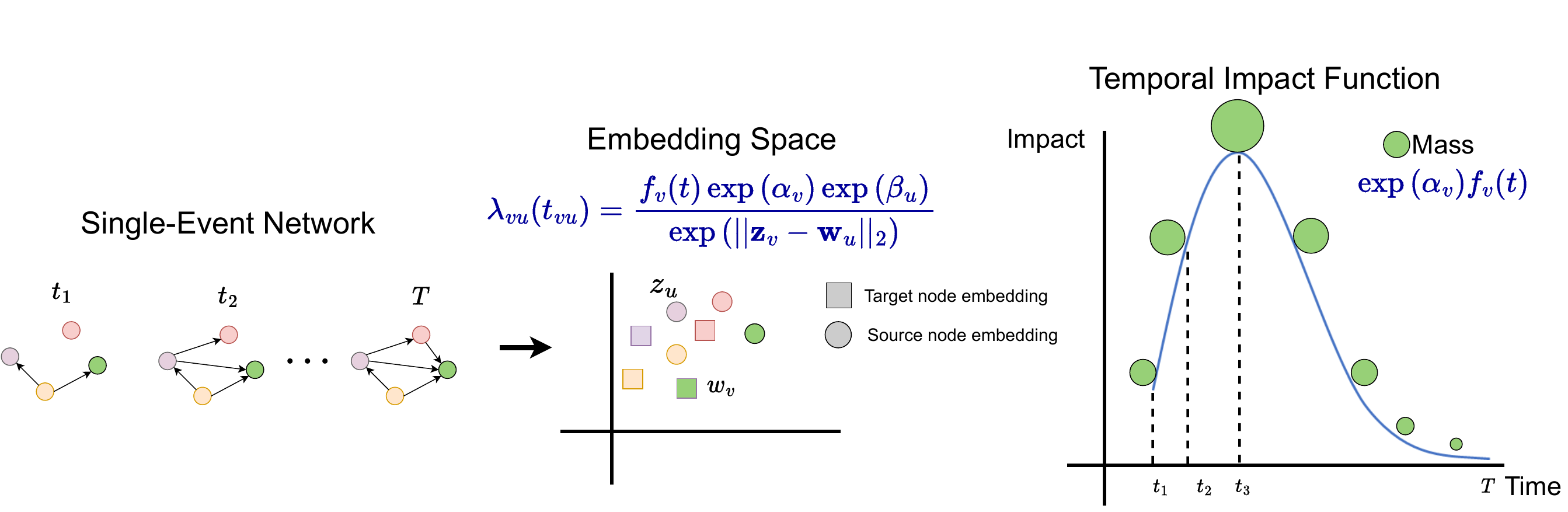}
\caption{\textsc{\modelabbrev} procedure overview. 
The model defines for the SE-PP an intensity function introducing two sets of static embeddings distinguishing between source $\mathbf{w}_u$ and target $\mathbf{z}_v$ node embeddings. Furthermore, each node is assigned its own random effect, distinguishing again the source $\beta_u$ and target $\alpha_v$ roles. The random effects can be parameterized to represent source and target masses through the exponential function. Finally, for each target node of the network, the model defines an impact function $f_v(t)$ yielding a temporal impact characterization of the nodes' link dynamics, which controls the nodes' time-varying mass as $\exp{(\alpha_v)}f_v(t)$.}\label{fig:\modelabbrev}
\end{figure*} 
Before presenting our modeling strategy for the links of networks, we will first establish the notations used throughout the paper. We utilize the conventional symbol, $\mathcal{G}=(\mathcal{V},\mathcal{E})$, to denote a directed Single-Event-Network over the timeline $[0,T]$ where $\mathcal{V}=\{1,\ldots,N\}$ is the vertex and $\mathcal{E} \subseteq \mathcal{V}^2\times[0,T]$ is the edge set such that each node pair has at most one link. Hence, a tuple, $(i,j,t_{ij}) \in\mathcal{E}$, shows a directed event (i.e., instantaneous link) from source node $j$ to target $i$ at time $t_{ij}\in[T]$, and there can be at most one $(i,j,t_{ij})$ element for each $(i,j)\in\mathcal{V}^2$ and some $t_{ij}\in[0,T]$. 

We always assume that the timeline starts at $0$ and the last time point is $T$, and we represent the interval by symbol, $[T]$. We employ $t_1\leq t_2\leq\cdots \leq t_N$  to indicate the appearance times of the corresponding nodes $1,2,\ldots,N\in\mathcal{V}$, and we suppose that node labels are sorted with respect to the minimum of their incoming edge times. In other words, if $i < j$, then we know that there exists $k\in\mathcal{V}$ such that $t_{ik} \leq t_{jl}$ $\forall (j,l,t_{jl}) \in \mathcal{E}$. 

The Inhomogeneous Poisson Point (IPP) process is a widely employed approach for modeling the number of events exhibiting varying characteristics depending on the time they occur \citep{GONZALEZ2016505}. They are parametrized by an \textit{intensity} or \textit{rate function} representing the average event density, such that the probability of sampling $m$ event points on the interval $[T]$ is
\begin{align}\label{eq:ipp_sampling_prob}
p_{M}(M(T)=m) := \frac{ \left[\Lambda(T)\right]^m }{m!}\exp(-\Lambda(T)),
\end{align}
where $M(T)$ is the random variable showing the number of events occurring over the interval $[T]$, and $\Lambda(T)$ is defined as $\int_{0}^{T}\lambda(t^{\prime})\mathrm{d}t^{\prime}$ for the intensity function $\lambda:[T]\rightarrow\mathbb{R}^+$. We refer unfamiliar readers to the work \citep{streit2010poisson} for more details concerning the process.

We employ a Poisson point process for characterizing the occurrence time of a link (i.e., a single event point indicating the publication or citation time), unlike their conventional practice in modeling the occurrence of an arbitrary number of events between a pair of nodes. 
Hence, we suppose that a pair can have at most one interaction (i.e., link), and we discretize the probability of sampling $m$ events given in Eq. \eqref{eq:ipp_sampling_prob} as having an event and no event cases. More formally, by applying Bayes' rule, we can write it as a conditional distribution of $M(t)$ being equal to $m \in \{0,1\}$ as follows:
\begin{align}
p_{M|M\leq 1} ( M(T) =  m ) = 
 \frac{ p_M\left( M(T) = m \right) }{ \sum_{m^\prime=0}^1 p_M\left( M(T) = m^\prime \right)} 
\nonumber\\
= \frac{ \exp\left(-\Lambda(T)\right)\left[\Lambda(T)\right]^m }{ \exp(-\Lambda(T)) + \exp(-\Lambda(T))\Lambda(T) }.
\end{align}
Therefore, the conditional probability of an event for the proposed \textit{Single-Event Poisson Process} is equal to:
\begin{equation}\label{eq:sepp_probability}
p_{M|M\leq 1}\left( M(T) = 1 \right) = \frac{\Lambda(T)}{1 + \Lambda(T)}.
\end{equation}
It is also not difficult to derive the likelihood function of the process based on Eq \eqref{eq:sepp_probability}. Let $(Y,\Theta)$ be random variables where $Y$ shows whether a link exists and $\Theta$ indicates the time of the corresponding link (if it exists). Then, we can write the likelihood of $(Y,\Theta)$ evaluated at $(1, t^{*})$ as follows:
\begin{align}\label{eq:prob_single_event}
{p_{Y,\Theta}\left(1,t^{*}\right)} &= p_Y\left\{ {Y = 1} \right\} p_{\Theta|Y}\left\{ \Theta=t^{*} | Y=1  \right\} \nonumber\\
&= \left(\frac{\Lambda(T)}{1+\Lambda(T)}\right) \left(\frac{ \lambda(t^{*}) }{ \Lambda(T) }\right)= \frac{ \lambda(t^{*}) }{1+\Lambda(T)}.
\end{align}

As a result, we can write the log-likelihood of the whole network by assuming that each dyad follows the Single-Event Poisson Process as follows:
\begin{align}\label{eq:ipp_likelihood}
&\mathcal{L}_{SE-PP}  (\Omega) := \log p( \mathcal{G} | \Omega )\nonumber \\ 
&= \sum_{1\leq i,j\leq N}\Big( y_{ij}\log\lambda(t_{ij}) - \log\big( 1 + \Lambda_{ij}(t_i,T) \big) \Big),
\end{align}

where $\Omega$ is the model hyper-parameters and $\Lambda_{ij}(t_i,T) := \int_{t_i}^T\lambda_{ij}(t^{\prime})\mathrm{d}t^{\prime}$. Note that for a homogeneous Poisson process with constant intensity $\lambda_{ij}$ for each $(i,j)$ pair, the probability of having an event throughout the timeline is equal to $\Lambda_{ij}(T) / (1 + \Lambda_{ij}(T)) = T\lambda_{ij} / (1 + T\lambda_{ij})$ by Eq. \eqref{eq:sepp_probability}. In this regard, the objective function stated in Eq. \eqref{eq:ipp_likelihood} is equivalent to a static Bernoulli model \citep{Hoff}:
\begin{align}\label{eq:bern_likelihood}
\mathcal{L}_{Bern}(\Omega) &:= \log p(\mathcal{G}|\Omega)\nonumber \\
&= \sum_{\substack{i,j\in\mathcal{V}}}\Big(y_{ij}\log{(\Tilde{\lambda}_{ij})}- \log\left(1+\!\Tilde{\lambda}_{ij}\right) \Big),
\end{align}
using the re-parameterization $\Tilde{\lambda}_{ij} := T\lambda_{ij}$.

\textbf{Comparison with a Bernoulli process DISEE:} The reason for not working with the Bernoulli Process, is its discrete nature, modeling independent trials indexed by a set and not a time variable \citep{grim}. The process is usually defined at discrete time points (e.g., coin tosses every minute). Contrary, the Poisson Process is a continuous counting process that models the number of events occurring over continuous intervals of time.

\section{Dynamic Impact Characterization}\label{sec:DIC}

In the realm of impact analysis and risk assessment, characterizing dynamic events is pivotal in understanding and managing potential consequences. We know that papers generally undergo the process of aging over time since novel works introduce more original concepts. we model the distribution of the impact of paper $i\in\mathcal{V}$ by the probability density function (pdf) of the \textsc{Log Normal} distribution:

\begin{equation}
f_i(t) = \frac{1}{t\sigma\sqrt{2\pi}}\exp\left( - \frac{ (\ln{t}-\mu)^2}{ 2\sigma^2 } \right)
\end{equation}
where $\mu$ and $\sigma$ are the parameters of the distribution.


In addition, as an alternative impact function, we consider the \textsc{Truncated} normal distribution:

\begin{equation}
f_i(t) = \frac{1}{\sigma}\frac{\phi(\frac{t-\mu}{\sigma})}{\Phi(\frac{\kappa-\mu}{\sigma})-\Phi(\frac{\rho-\mu}{\sigma})}
\end{equation}

where $\mu$ and $\sigma$ are the parameters of the distribution which lie in $(\rho,\kappa) \in \mathbb{R}$, $\phi(x)=\frac{1}{\sqrt{2\pi}}\exp{(-\frac{1}{2}x^2)}$, and $\Phi(\cdot)$ is the cumulative distribution function (cdf) $\Phi(x)=\frac{1}{2}\Big(1+ \text{erf}(\frac{x}{\sqrt{2}})\Big)$. Such distributions are particularly valuable for capturing the inherent variability and asymmetry in the life-cycle of a paper.

As argued in the work of \cite{ts5} employing an impact function based solely on the pdf yields a successful characterization of the aging of a paper but does not account for the preferential attachment connecting to papers regarding their impact at a specific time moment. For that, we also consider the approach in \cite{ts5} as an impact function, which is the product of the pdf and the cumulative distribution function (cdf), as:
\begin{equation}\label{eq:paag}
f_i(t) = pdf_i(t)cdf_i(t),
\end{equation}

We here note that Eq. \eqref{eq:ipp_likelihood} requires the calculation of the integral $\Lambda_{ij}(t_i,T) := \int_{t_i}^T\lambda_{ij}(t^{\prime})\mathrm{d}t^{\prime}$ which can be trivially calculated for an impact function based on Eq. \eqref{eq:paag}, by 
applying the substitution rule for $u:=cdf_i(t)$. It is also important that the cdf is of analytical form which is the case for both the \textsc{Truncated} normal and \textsc{Log-Normal} distributions we consider.

\section{Single-Event Network Embedding by the Latent Distance Model}\label{sec:DISEE}
Our main purpose is to represent every node of a given single-event network in a low $D$-dimensional latent space ($D \ll N$) in which the pairwise distances in the embedding space should reflect various structural properties of the network, like homophily and transitivity \citep{nakis22hbdm,HMLDM}. For instance, in the \textit{Latent Distance Model} \citep{Hoff}, one of the pioneering works, the probability of a link between a pair of nodes depended on the log-odds expression, $\gamma_{ij}$, as $\alpha - \| \mathbf{z}_i -\mathbf{z}_j \|_2$ where $\{\mathbf{z}_i\}_{i\in\mathcal{V}}$ are the node embeddings, and $\alpha\in\mathbb{R}$ is the global bias term responsible for capturing the global information in the network. It has been proposed for undirected graphs but can be extended for directed networks as well by simply introducing another node representation vector $\{\mathbf{w}_i\}_{i\in\mathcal{V}}$ in order to differentiate the roles of the node as source (i.e., sender) and target (i.e., receiver). By the further inclusion of two sets of random effects $(\alpha_i,\beta_i)_{i\in\mathcal{V}}$ describing the in and out degree heterogeneity, respectively, we can define the log-odds (Bernoulli)  and log-rate (Poisson) \citep{nakis22hbdm}  expression as:
\begin{equation}\label{eq:natural_param_defn}
\gamma_{ij}=\alpha_i+\beta_j - \| \mathbf{z}_i - \mathbf{w}_j \|_2.
\end{equation}
We can now combine a dynamic impact characterization function with the \textit{Latent Distance Model}, to obtain an expression for the intensity function of the proposed \textit{Single-Event Poisson Process}, as:
\begin{equation}
\label{eq_ldm_in}
    \lambda_{ij}(t_{ij})=\frac{f_i(t_{ij})\exp{(\alpha_i)}\exp{(\beta_j)}}{\exp{(\| \mathbf{z}_i - \mathbf{w}_j \|_2)}}.
\end{equation}

Combining the intensity function of Eq. \eqref{eq_ldm_in} with the log-likelihood expression of Eq. \eqref{eq:ipp_likelihood} yields the \textit{\modelname} (\textsc{\modelabbrev}). Under such a formulation, we exploit the time information data indicating when links occur through time, so we can grasp a more detailed understanding of the evolution of networks, generate enriched node representations, and quantify a node's temporal impact on the network.

\textbf{Generative Model}: To generate Single-Event Networks defining accurate and powerful impact functions, we consider the following generative process:
\begin{enumerate}
    \item Input: $\alpha_{\lambda}, \theta_{\lambda}, \alpha_{\beta}, \theta_{\beta}, \mu_m, \sigma_m, \alpha_s, \theta_s, \sigma_{\mathbf{z}}, \sigma_{\mathbf{w}}$
    \item For each node $i \in \mathcal{V}$
    \begin{enumerate}
        \item $t_i \sim Uniform(0,T)$, \hfill Paper appearance times
        \item $\lambda_i \sim Gamma(\alpha_{\lambda}, \theta_{\lambda})$, \hfill Impact cited paper
        \item $\beta_i \sim Gamma(\alpha_{\beta},\theta_{\beta})$, \hfill Degree citing paper
        \item $\mathbf{z}_i \sim \mathcal{N}(\bm{0}, \sigma_{\mathbf{z}}^2\mathbf{I})$, \hfill Cited paper embedding
        \item $\mathbf{w}_i \sim \mathcal{N}(\bm{0}, \sigma_{\mathbf{w}}^2\mathbf{I})$, \hfill Citing paper embedding
        \item $\mu_i \sim \mathcal{N}(\mu_m, \sigma_m^2)$, \hfill Param. of log-normal 
         \item $\sigma_i \sim Gamma(\alpha_s, \theta_s)$, \hfill Param. of log-normal 
        \item $f_i  \leftarrow \text{ pdf of } LogNormal(\mu_i, \sigma_i)$ 
    \end{enumerate}
    \item Relabel nodes $(1,\ldots,N)$ such that $t_1\leq \cdots \leq t_N$
    \item For each node $i \in \mathcal{V}$ 
    \begin{enumerate}
        \item $\kappa_i \sim Poisson(\lambda_i)$, \hfill Number of citations
        \item For each node $j > i$: \hfill $p_{ij} \leftarrow \frac{ \left(\kappa_if_i(t_j-t_i)\right) \beta_j }{ \exp{\left(\|\mathbf{z}_i - \mathbf{w}_j \|\right)} }$
        \item $\mathbf{p} \leftarrow (p_{i(i+1)},\ldots,p_{ij},\ldots,p_{iN})/\sum_{k=i+1}^Np_{ik}$
        \item $K \leftarrow \min(\kappa_i, N-i)$\\
         $(j_1,\ldots,j_K) \sim Mult(\mathbf{p}, \text{sample }K\text{ w.o. repl.})$ 
         \item Add links $(i, i+j_1),\ldots, (i,i+j_K)$                
    \end{enumerate}
\end{enumerate}
We could, in Step 4, have generated the observed networks from the Single-Event Poisson Process (SE-PP); however, this would technically assign separate time points to be sampled for each citation. Instead, we draw the citations from fixed time points of publications using multinomial sampling without replacement.

\textbf{Model ablations}: We define an Impact Function Model \textsc{(IFM)}, where only the impact function is fitted to the target nodes (cited papers) describing their link (citation) dynamics. Comparing with \textsc{IFM} will allow us to validate the quality of the impact characterization of \textsc{\modelabbrev}. We further contrast our model to a Preferential Attachment Model (\textsc{PAM}) setting where the embedding dimension is set as $D=0$, providing a quantification of the importance of including an impact function and an embedding space in \textsc{\modelabbrev}. In addition, we consider a combination of an Impact Function Model with a Preferential Attachment Model, defining a Temporal Preferential Attachment Model (\textsc{TPAM}). Compared with the \textsc{TPAM} we aim to verify the importance of introducing an embedding space characterization in citation networks. Finally, we systematically contrast the performance of \textsc{\modelabbrev} to conventional static latent distance modeling \textsc(LDM) \citep{Hoff} corresponding to setting the impact function to be constant $f_{i}(t)\propto 1$ in \textsc{\modelabbrev}. The \textsc{LDM} is a very powerful link predictor \citep{nakis2022a,HMLDM} and contrasting its performance against \textsc{\modelabbrev} will help us showcase the successful reconciliation of static latent space network embedding approaches with classical dynamic impact assessments of citation networks. In Table \ref{tab:abl}, we provide the rate formulation of each considered model ablations and the corresponding model characteristics in terms of impact characterization and definition of an embedding space.

\begin{table*}
    \centering
    \begin{minipage}[b]{0.56\textwidth}
        \centering
        \caption{\textsc{\modelabbrev} model and considered model ablations.}
        \label{tab:abl}
        \resizebox{1.2\textwidth}{!}{%
            \begin{tabular}{l|ccccc}
                \toprule
                \textbf{Model Name} & \textsc{IFM} & \textsc{PAM} & \textsc{TPAM} & \textsc{LDM} & \textsc{DISEE} \\
                \midrule
                \textbf{Rate formulation} & $\exp{(\alpha_i)}f_i(t)$ & $\exp{(\alpha_i)}\exp{(\beta_j)}$ & $f_i(t)\exp{(\alpha_i)}\exp{(\beta_j)}$ & $\frac{\exp{(\alpha_i)}\exp{(\beta_j)}}{\exp{(||\bm{z}_i-\bm{w}_j||_2)}}$ & $ \frac{f_i(t)\exp{(\alpha_i)}\exp{(\beta_j)}}{\exp{(||\bm{z}_i-\bm{w}_j||_2)}}$ \\
                \textbf{Impact} & \cmark & \xmark & \cmark & \xmark & \cmark \\
                \textbf{Embedding space} & \xmark & \xmark & \xmark & \cmark & \cmark \\
                \bottomrule
            \end{tabular}
        }
    \end{minipage}
    \hfill
    \begin{minipage}[b]{0.35\textwidth}
        \centering
        \caption{Statistics of networks.}
        \label{tab:dataset_statistics}
                \resizebox{0.8\textwidth}{!}{%
        \begin{tabular}{@{}lccc@{}}
            \toprule
             & $\left|\mathcal{V}_1\right|$ & $\left|\mathcal{V}_2\right|$ & $\left|\mathcal{E}\right|$  \\ 
             \midrule
             \textsl{Artificial} (\textsl{Art}) & 4,001 & 4,979 & 160,039 \\
            \textsl{Machine Learning} (\textsl{ML}) & 22,540 & 148,703 & 526,226 \\
            \textsl{Physics} (\textsl{Phys}) & 20,012 & 51,996 & 573,378  \\
            \textsl{Social Science} (\textsl{SoSci}) & 12,930 & 100,402 & 288,012  \\
            \bottomrule
        \end{tabular}%
        }
    \end{minipage}
\end{table*}

\section{Results and Discussion}\label{sec:results}
In this section, we will evaluate how successfully \textsc{\modelabbrev} reconciles traditional impact quantification approaches with latent distance modeling. We denote as \textsc{\modelabbrev} the model with an impact function given as a probability density function ($f_i(t)=pdf_i(t)$), and as \textsc{\modelabbrev\ PA} the Preferential Attachment (PA) model with an impact function based on Eq. \eqref{eq:paag}. Lastly, we also consider the Fixed Impact (FI) variants of the models, where the impact functions are fixed, during training, to the empirical distribution of the citation data distributions, yielding the \textsc{FI-\modelabbrev} and \textsc{FI-\modelabbrev\ PA} models. The latter two models essentially combine an \textsc{IFM} model with the proposed \textsc{\modelabbrev} model. All experiments regarding the \textsc{DISEE} models have been conducted on an $8$ GB NVIDIA RTX $2070$ Super GPU. In addition, we adopted the Adam optimizer \citep{kingma2017adam} with a learning rate of $\text{lr}=0.1$ and ran it for $3000$ iterations. Our objective was to minimize the negative log-likelihood of the Single-Event Poisson Process, utilizing a case-control approach \citep{case_control} to achieve scalable inference. (\textit{For extensive training and implementation details, code release, as well as, a complexity analysis please visit the supplementary.})

We test the proposed approach's effectiveness in the link prediction task by comparing it to the classical LDM which is not time-aware and does not quantify temporal impact, as well as, against four prominent baselines. We also consider multiple model ablations that are either able to characterize a node's impact or account for \textsc{GRL}, i.e. define node embeddings, but not both. For the task of link prediction, we remove $20\%$ of network links and we sample an equal amount of non-edges as negative samples and construct the test set. 
The link removal is designed in such a way that the residual network stays connected. 
For the evaluation, we consider the Precision-Recall Area Under Curve (AUC-PR) score, as it is a metric not sensitive to the class imbalance between links and non-links (\textit{Receiver Operator Characteristic scores are provided in the supplementary}). We then continue by evaluating the quality of impact expression of \textsc{\modelabbrev} by visually presenting the inferred impact functions and comparing them against the \textsc{FI-\modelabbrev} model. Finally, we visualize the model's learned temporal space representing the target papers, accounting for their temporal impact in terms of their mass at a specific time point, and characterizing the different papers' lifespans.

\textbf{Datasets}: In our experiments, we employ three real citation networks. Specifically, we use the OpenAlex dataset \citep{priem2022openalex}, exploring highly impactful scientific domains such as (i) \textit{Machine Learning}, (ii) \textit{Physics}, and (iii) \textit{Social Science}. Lastly, we also consider an artificial network generated based on the generative process as described in Section \ref{sec:DISEE}. In order to be able to characterize scientific impact, we consider papers that have been cited at least ten times while we assign zero mass to papers with less than ten incoming links. 
This yields a directed bipartite structure where target nodes have at least ten citations. Analytically, the network statistics are given in Table \ref{tab:dataset_statistics} where $|\mathcal{V}_1|$ is the number of target nodes, $|\mathcal{V}_2|$ the number of source nodes, and $|\mathcal{E}|$: the total number of links. (\textit{For more details see supplementary.})

\textbf{Baselines}: In addition to the model ablations stated in Table \ref{tab:abl}, we employed four baselines relying on different encoding strategies. (i) \textsc{Node2Vec} \citep{node2vec-kdd16} is a well-known random walk-based approach that learns node embeddings by utilizing the co-occurrence frequencies of nodes within the generated node sequences. (ii) \textsc{CTDNE} \citep{rw_3} can be considered as its adaptation for temporal networks. It performs random walks based on the temporal characteristics of the network thereby incorporating the time information of the links. (iii) \textsc{Verse} \citep{tsitsulin2018verse} aims to extract latent representations reflecting Personalized Page Rank (PPR) similarity among nodes. Finally, we have employed the (iv) \textsc{NetSMF} model \citep{qiu2019netsmf} which learns embeddings by decomposing a designed matrix representing the high and low-order node interactions. (\textit{For more details see supplementary.})

\begin{table*}[!t]
\caption{AUC-PR scores over four networks. (Bold numbers denote the best-performing model.)}
\label{tab:pr-bip}
\begin{center}
\resizebox{.96\textwidth}{!}{%
\begin{tabular}{lcccccccccccccccc}\toprule
\multicolumn{1}{l}{} & \multicolumn{4}{c}{\textsl{Art}} & \multicolumn{4}{c}{\textsl{ML}} & \multicolumn{4}{c}{\textsl{Phys}} & \multicolumn{4}{c}{\textsl{SoSci}}\\\cmidrule(rl){2-5}\cmidrule(rl){6-9}\cmidrule(rl){10-13}\cmidrule(rl){14-17}
\multicolumn{1}{r}{Dimension ($D$)}  &1    &2	&3  &8 &1    &2	&3  &8   &1    &2 	&3  &8   &1    &2  &3 &8 \\\cmidrule(rl){1-1}\cmidrule(rl){2-5}\cmidrule(rl){6-9}\cmidrule(rl){10-13}\cmidrule(rl){14-17}
\textsc{IFM}    & \multicolumn{4}{c}{.882} & \multicolumn{4}{c}{.610} & \multicolumn{4}{c}{.639} & \multicolumn{4}{c}{.572}\\
\textsc{PAM}    & \multicolumn{4}{c}{.520} & \multicolumn{4}{c}{.810} & \multicolumn{4}{c}{.838} & \multicolumn{4}{c}{.796}\\
\textsc{TPAM}   & \multicolumn{4}{c}{.903} & \multicolumn{4}{c}{.806} & \multicolumn{4}{c}{.836} & \multicolumn{4}{c}{.790}\\
\textsc{Node2Vec} &.593   &.810	&.767  &.828    &.584   &.799	&.873  &.966    &.592    &.811	&.896	&.974    &.566    &.800  &.897  & .964   \\
\textsc{Verse}   &.606   &.816	&.774  & .831   & .760  &.911	&.946  & .981   & .780   &.882	&.929	& .977   &.800    & .919 &.949  &\textbf{.976}    \\
\textsc{CTDNE}   & .651  &.711	&.768  &.810    &.653   &.768	& .830 &.928    & .636   &.777	&.835	& .937   & .677   & .805 &.859  &.924    \\
\textsc{NetSMF} &.749   &.834	&.834  & .847   & .713  &.750	& .818 & .913   &.750    &.790	&.867	&.936    &  .772  & .836 &.836  & .923   \\
\textsc{LDM}     &.943  & .953  & .958  &.955   & \textbf{.953}    &.969	& .976 & \textbf{.983}   &\textbf{.940}    &\textbf{.963} 	&\textbf{.973}	& \textbf{.987}   &\textbf{.943}   &.956  & 0.963 & .975   \\
\midrule
\textsc{\modelabbrev \: PA }   &.947  &.955   &.957   & .956  & .944   &.967	&.973  & .979   &.926    &.958	&.968	&.981    &.930    &.958  &.965  & .971   \\
\textsc{FI-\modelabbrev \: PA }    &.944  & .954  & .956  &.956   & .941  &.966	&.973  &.978    &.923    &.956	&.967	&.980    & .929   &.956  & .963 &.970    \\
\textsc{\modelabbrev }   &\textbf{.955}  & \textbf{.962}  &\textbf{.963}   &\textbf{.961}   & \textbf{.953}    &\textbf{.970}	&\textbf{.977}  & .982   &.937    &\textbf{.963}  &\textbf{.973} & .984   &.939    &\textbf{.962}   &.968  & \textbf{.976}  \\
\textsc{FI-\modelabbrev}   & \textbf{.955} &\textbf{.962}   & \textbf{.963}  & \textbf{.961}  & .952   &\textbf{.970}	&\textbf{.977}  & .982 &.934  &.962    &.972 & .983 & .940  & \textbf{.962}   &\textbf{.969}  &.975     \\
\bottomrule  
\end{tabular}%
}
\end{center}
\end{table*}

\begin{figure}[b]
    \centering
     \begin{subfigure}[t]{0.23\textwidth}
        \includegraphics[width=\textwidth]{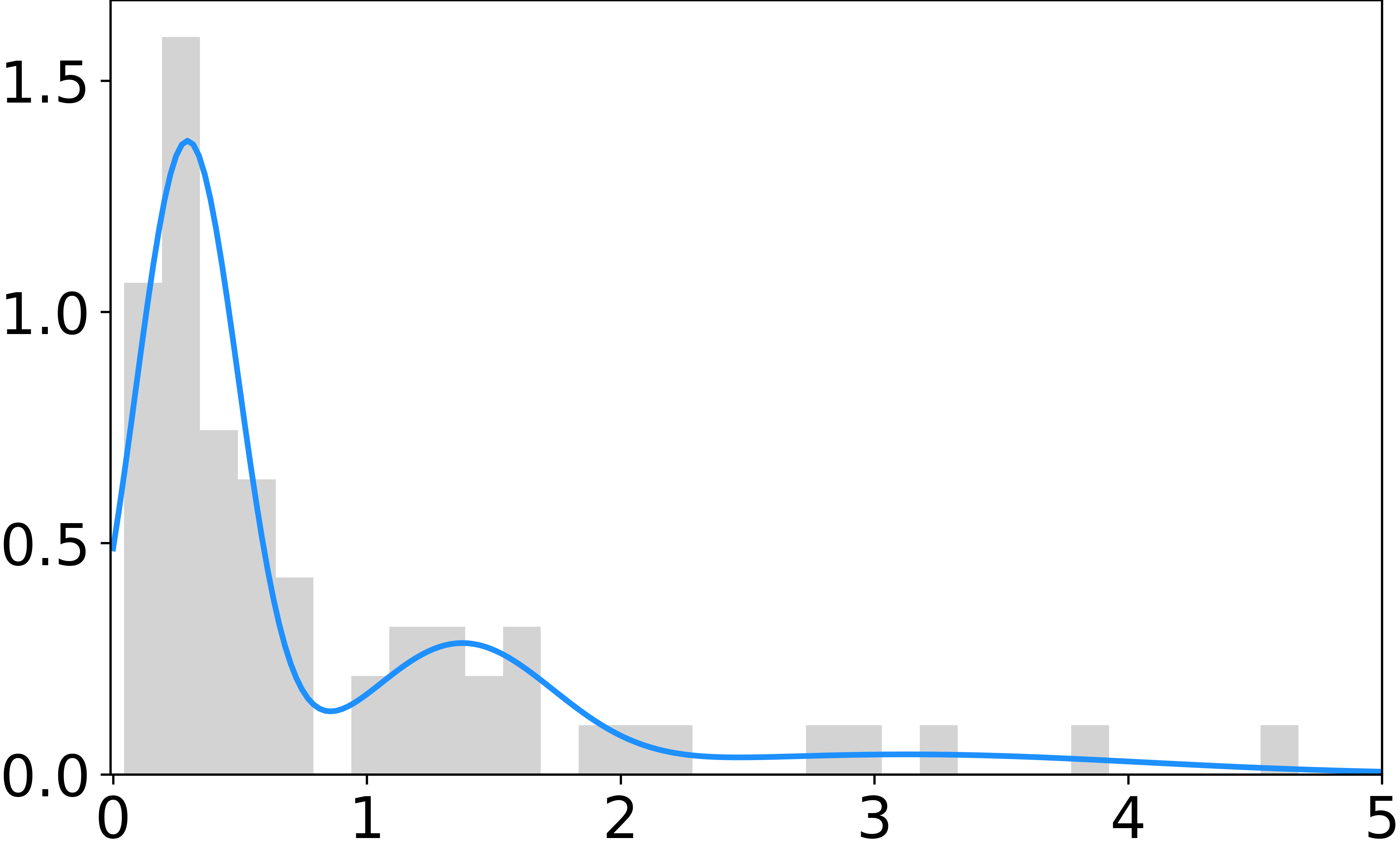}
        \caption{Artificial paper \# 1.}
    \end{subfigure}
    \hfill
    \begin{subfigure}[t]{0.23\textwidth}
        \includegraphics[width=\textwidth]{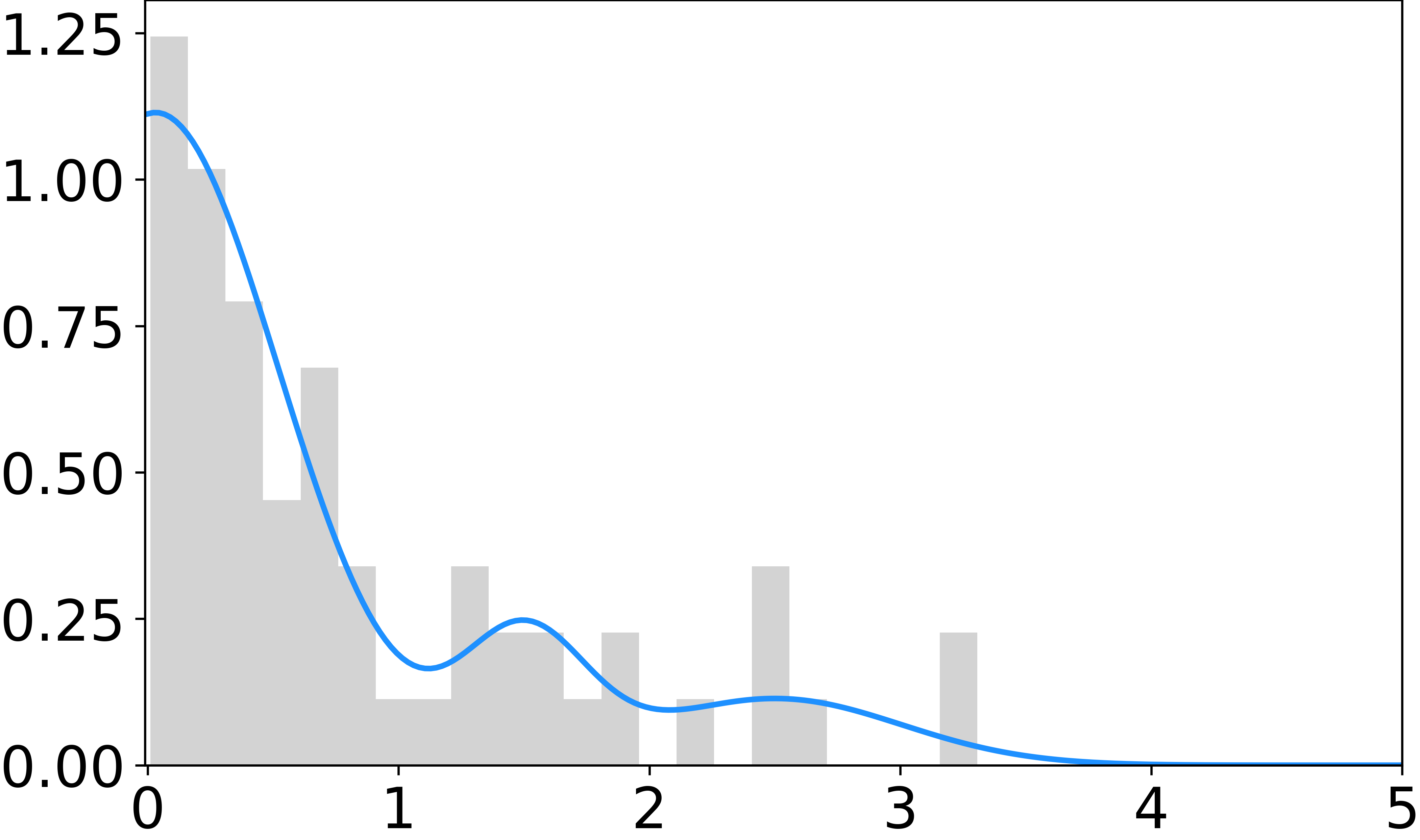}
        \caption{Artificial paper \# 2.}
    \end{subfigure}
    \caption{\textsl{Artificial}: 
    Comparison of the inferred impact functions, generated by a \textsc{\modelabbrev} Mixture Model with three \textsc{Truncated} normal distributions, to the true citation histogram of two papers from the \textsl{Art} dataset.}   
    \label{fig:mix_main}
\end{figure}
\textbf{Link prediction}: 
We here compare the results of \textsc{\modelabbrev} variants against the different model ablations and baselines in terms of their AUC-PR scores, as presented in Table \ref{tab:pr-bip}. Scores for our models are presented as the mean value of three independent runs of the Adam optimizer (error bars were found in the $10^{-4}$ scale and thus omitted). We here provide the scores for the \textsc{DISEE} model and variants under the \textsc{Log-Normal} distribution. Similar results are obtained while also considering the \textsc{Truncated} normal distribution, and they are provided in the supplementary. We here observe, that the best performance is achieved by model specifications that define an embedding space, i.e. the \textsc{DISEE} variants and \textsc{LDM} models. The Preferential Attachment Models in both the static (\textsc{PAM}) and temporal (\textsc{TPAM}) versions are characterized by an approximately $15\%$ decrease in their link prediction score. This highlights the importance and benefits of the predictive performance an embedding space provides. Contrasting the performance of \textsc{DISEE} against the \textsc{LDM}, we witness almost identical scores, verifying that \textsc{DISEE} successfully inherited the link prediction power of the \textsc{LDM}. Comparing now to the baselines, we here observe that all of the \textsc{DISEE} variants and the \textsc{LDM} are characterized by mostly a significantly higher performance. Notably, the \textsc{DISEE PA} and \textsc{FI-DISEE PA} models show a small decrease in the performance. Lastly, fixing the impact functions in the \textsc{FI-DISEE} and \textsc{FI-DISEE PA} models does not have an effect on the model performance. Contrasting the \textsc{DISEE} and \textsc{LDM} models on the \textsl{Art} network, we observe a higher performance from the former. However, this advantage diminishes as the dimensionality increases, attributed to the expression and conveyance of temporal information by the latent space.

\begin{figure*}[!t]
    \centering
     \begin{subfigure}[b]{0.24\textwidth}
        \includegraphics[width=\textwidth]{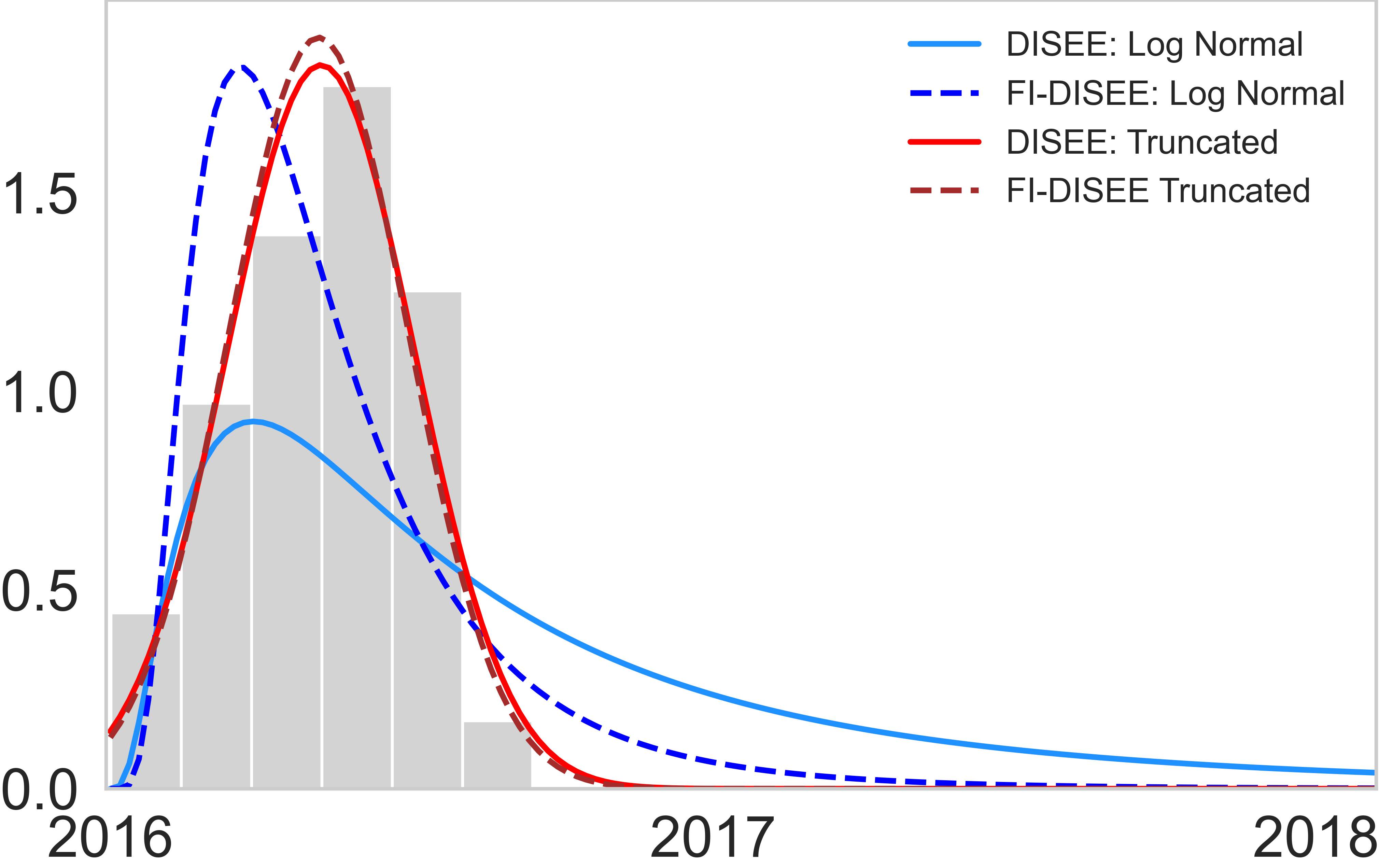}
        \caption{Deep Residual Learning for Image Recognition \citep{ifm1}.}
    \end{subfigure}
    \hfill
    \begin{subfigure}[b]{0.24\textwidth}
        \includegraphics[width=\textwidth]{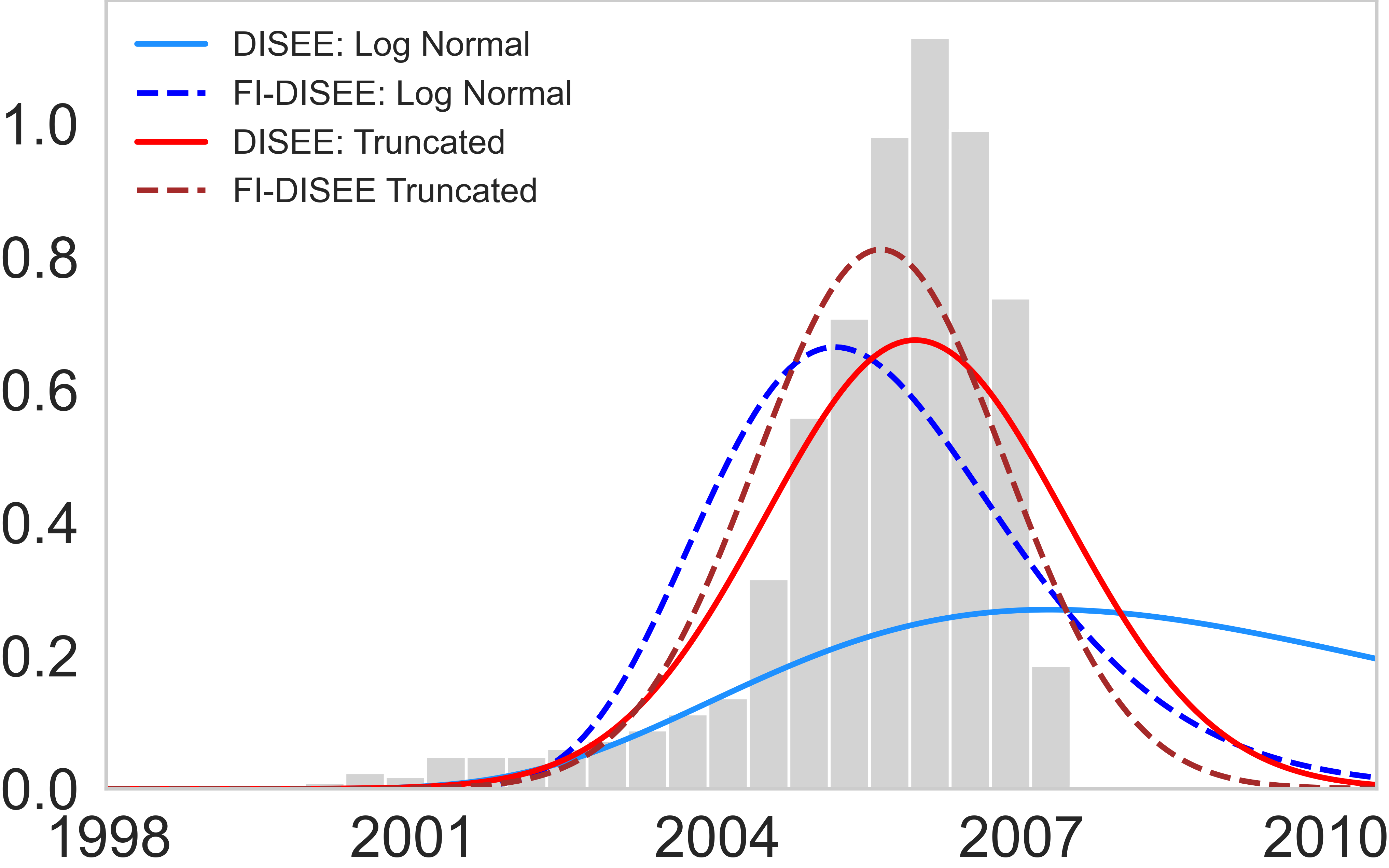}
        \caption{Gradient-based learning applied to document recognition \citep{ifm2}.}
    \end{subfigure}
    \hfill
    \begin{subfigure}[b]{0.24\textwidth}
        \includegraphics[width=\textwidth]{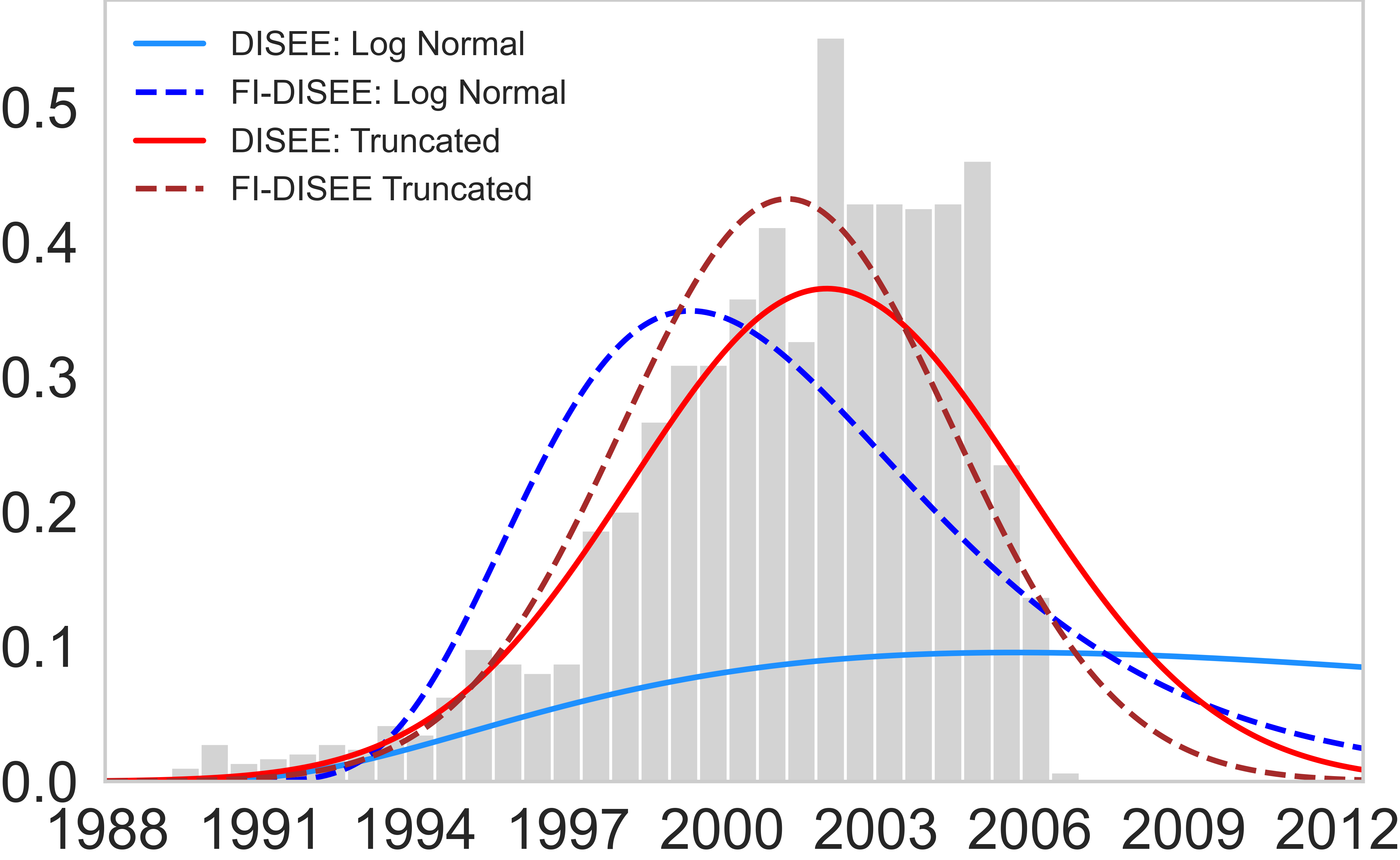}
        \caption{Structural equation modeling in practice \citep{ifm3}.}
    \end{subfigure}
    \hfill
    \begin{subfigure}[b]{0.24\textwidth}
\includegraphics[width=\textwidth]{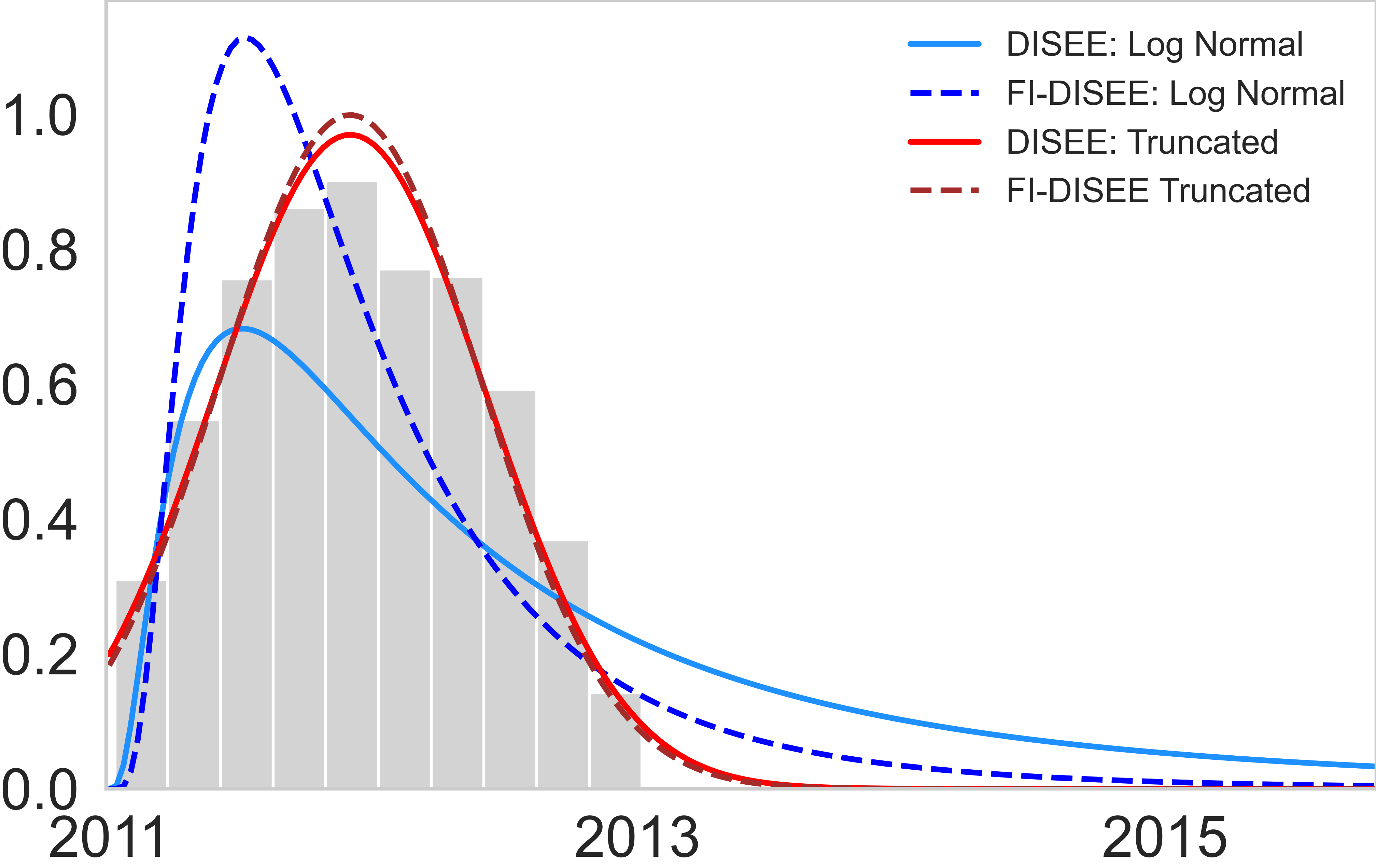}
        \caption{LIBSVM: A library for support vector machines \citep{ifm4}.}
    \end{subfigure}
    \caption{\textsl{Machine Learning}: \textsc{\modelabbrev} and \textsc{FI-\modelabbrev} models \textsc{Truncated} normal, and \textsc{Log-Normal} inferred impact function visualizations compared to the true citation histogram for four highly cited \textsl{ML} papers.}     
    \label{fig:ML_ln}
\end{figure*}

\begin{figure*}[!t]
\begin{subfigure}{0.24\textwidth}
        \includegraphics[width=1\textwidth]{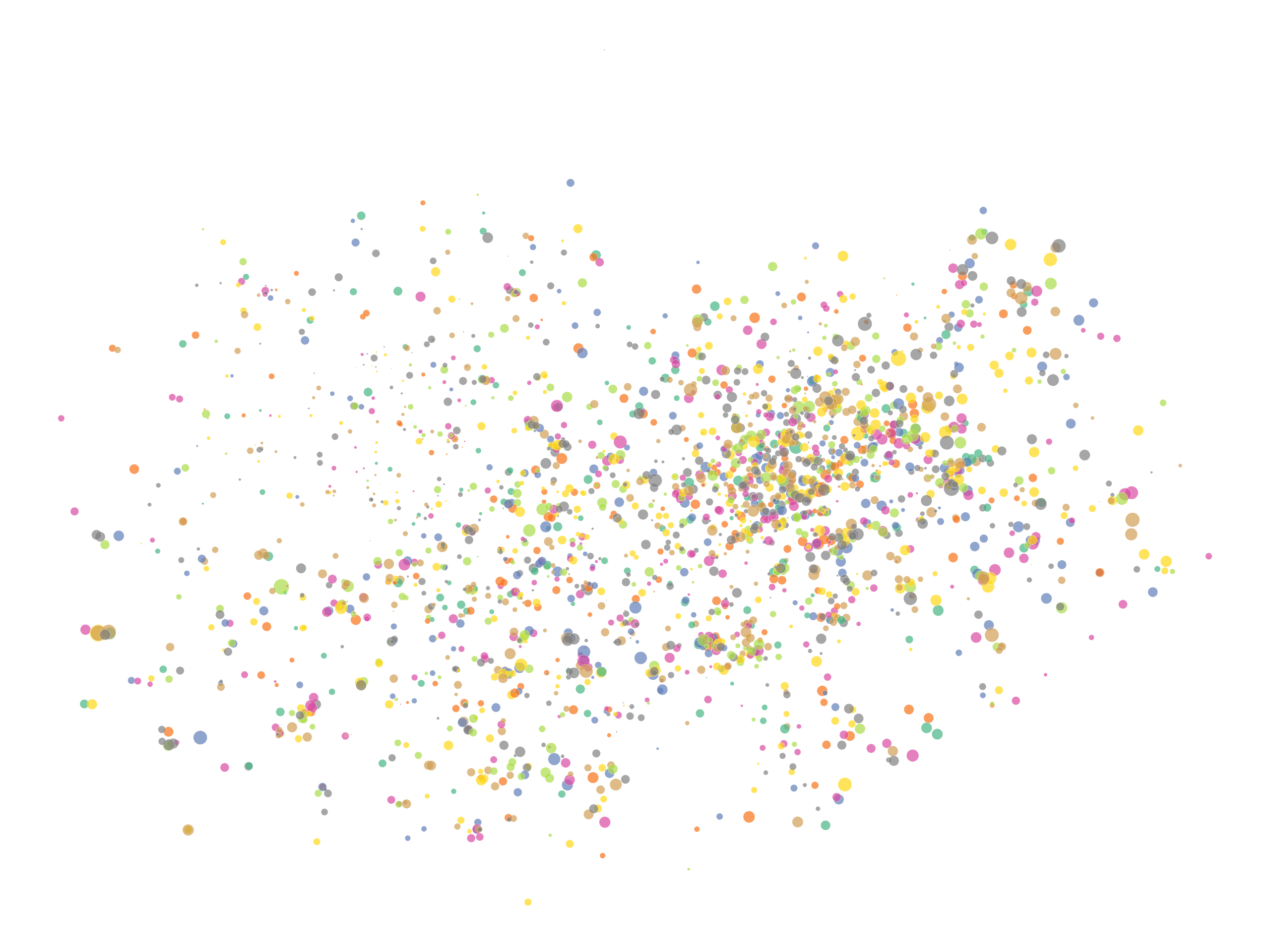}
        \caption{1998}
    \end{subfigure}
    \hfill
    \begin{subfigure}{0.24\textwidth}
        \includegraphics[width=1\textwidth]{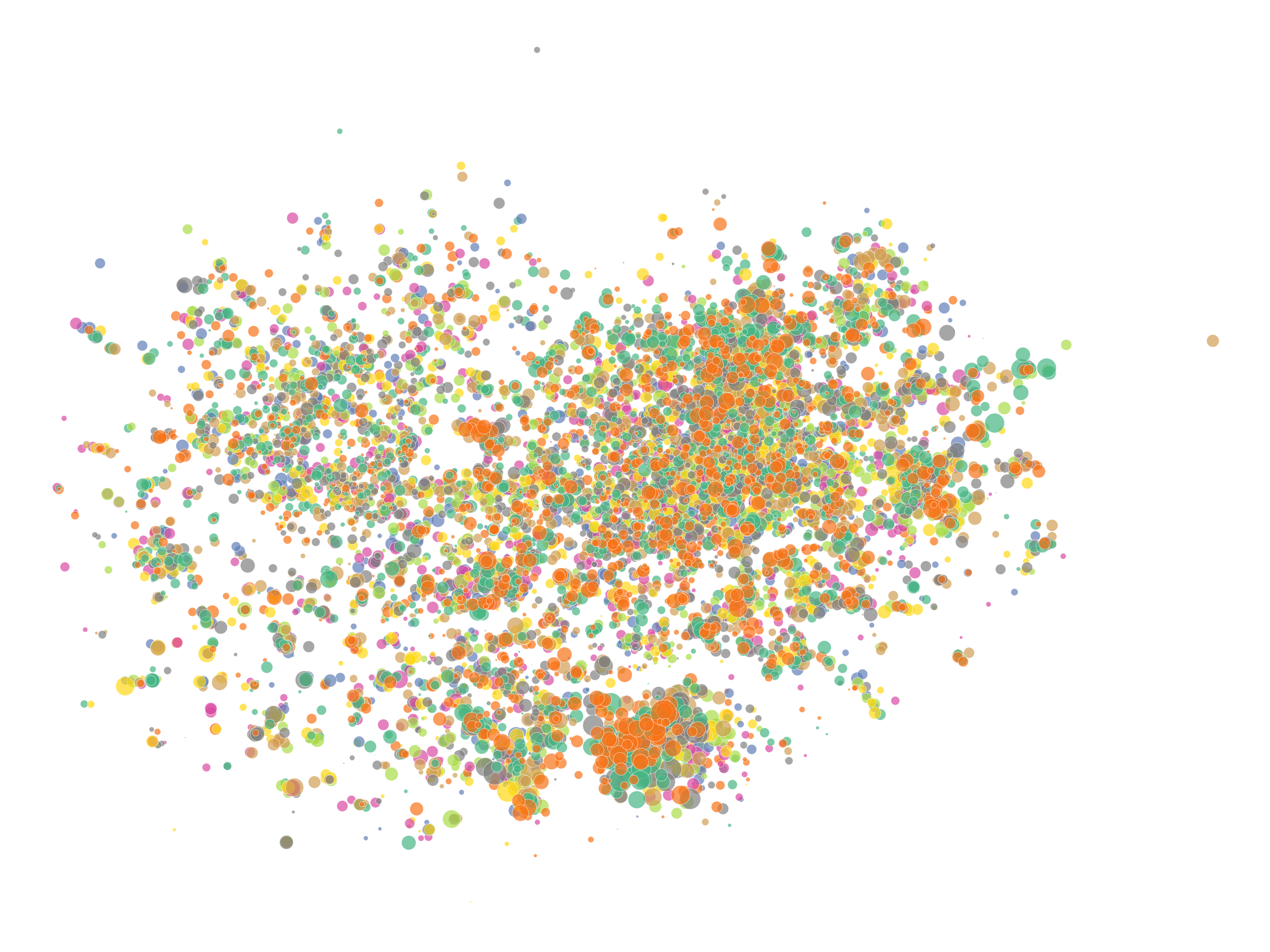}
        \caption{2008}
    \end{subfigure}
    \hfill
    \begin{subfigure}{0.24\textwidth}
        \includegraphics[width=1\textwidth]{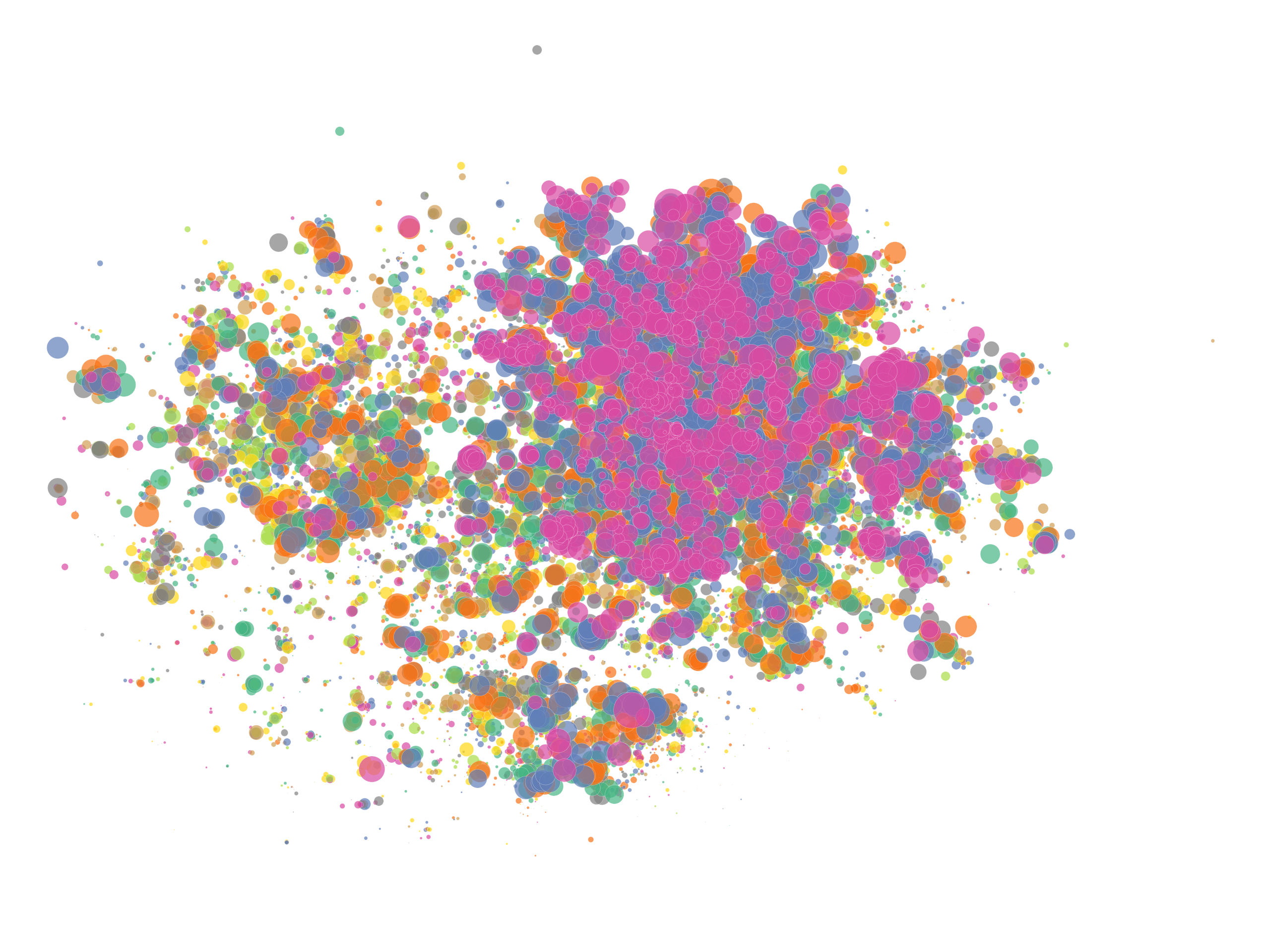}
        \caption{2018}
    \end{subfigure}
    \hfill
    \begin{subfigure}{0.24\textwidth}
        \includegraphics[width=1\textwidth]{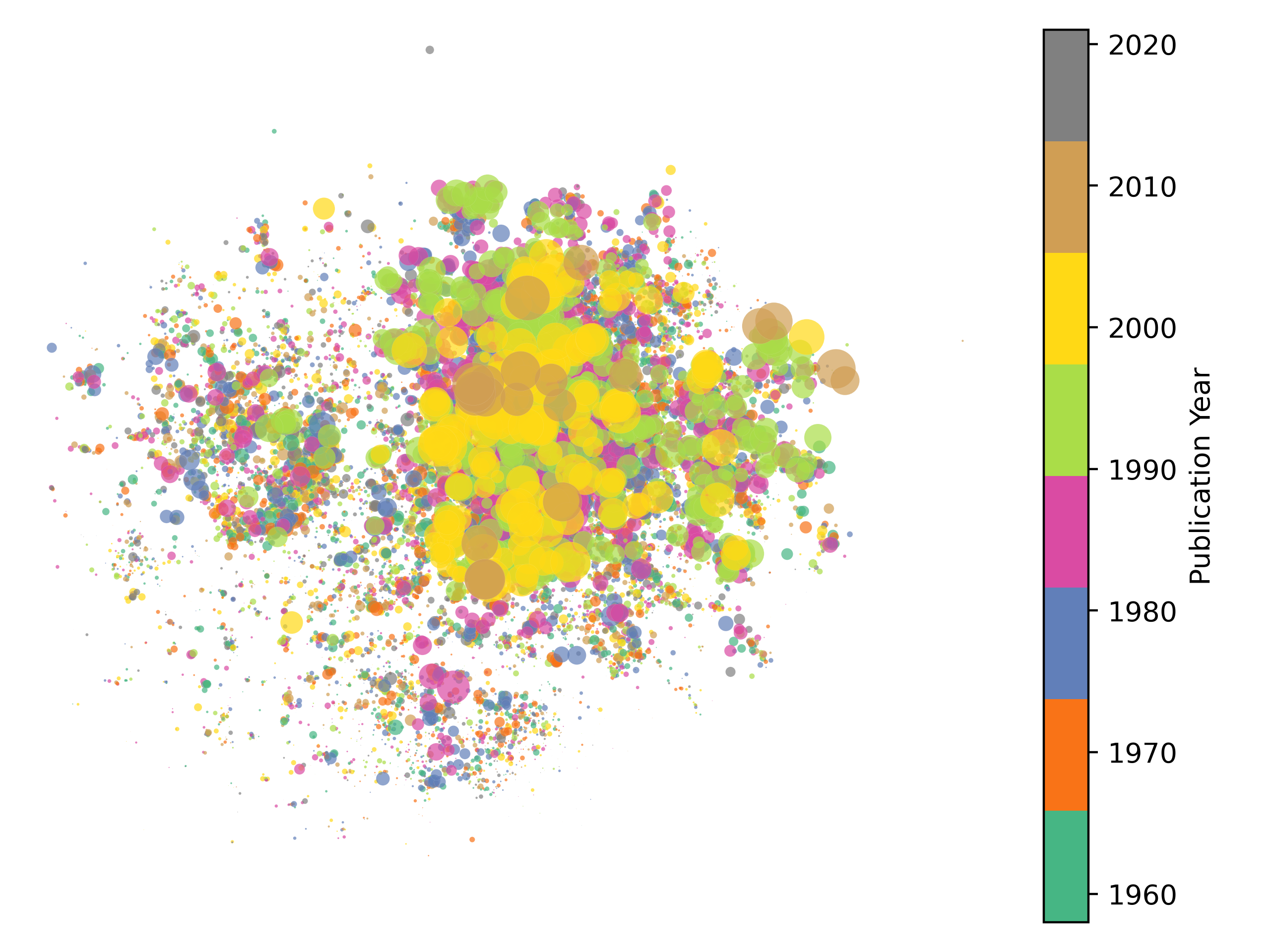}
        \caption{2023}
    \end{subfigure}
 \caption{\textsl{Machine Learning}: \textsc{\modelabbrev} 2-dimensional embedding space \textsc{Log-Normal} yearly evolution. Node sizes are based on each paper's mass, $f_i(t)\exp{(\alpha_i)}$. Nodes are color-coded based on their publication year.}
 \label{fig:history}
\end{figure*}

\begin{figure*}[!ht]
  \centering    
  \begin{subfigure}[t]{0.24\linewidth}
\includegraphics[width=\linewidth]{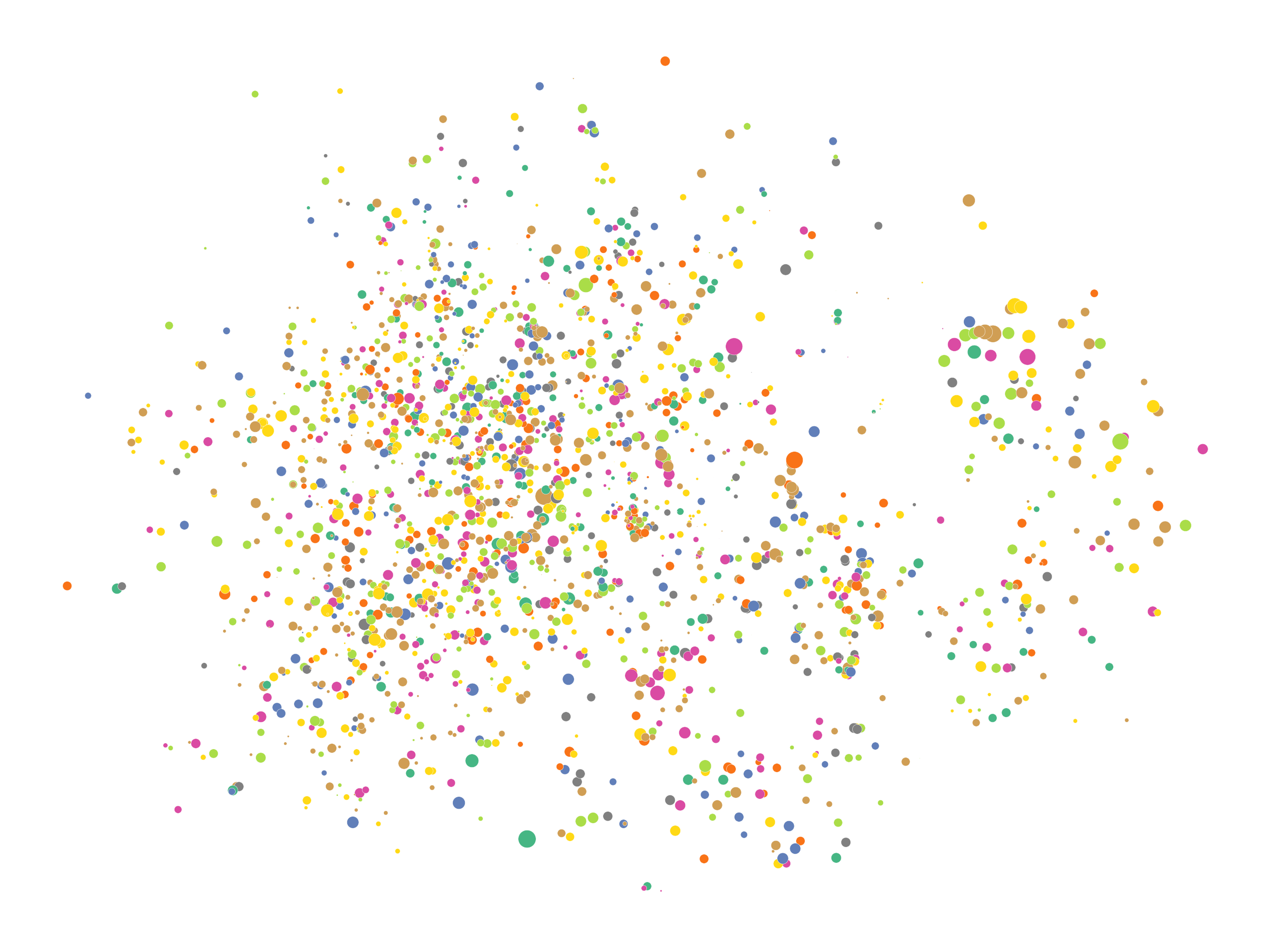}
    \caption{\textsl{SoSci} 1998}
    \label{fig:soc1998}
  \end{subfigure}
  \hfill
  \begin{subfigure}[t]{0.24\linewidth}
    \includegraphics[width=\linewidth]{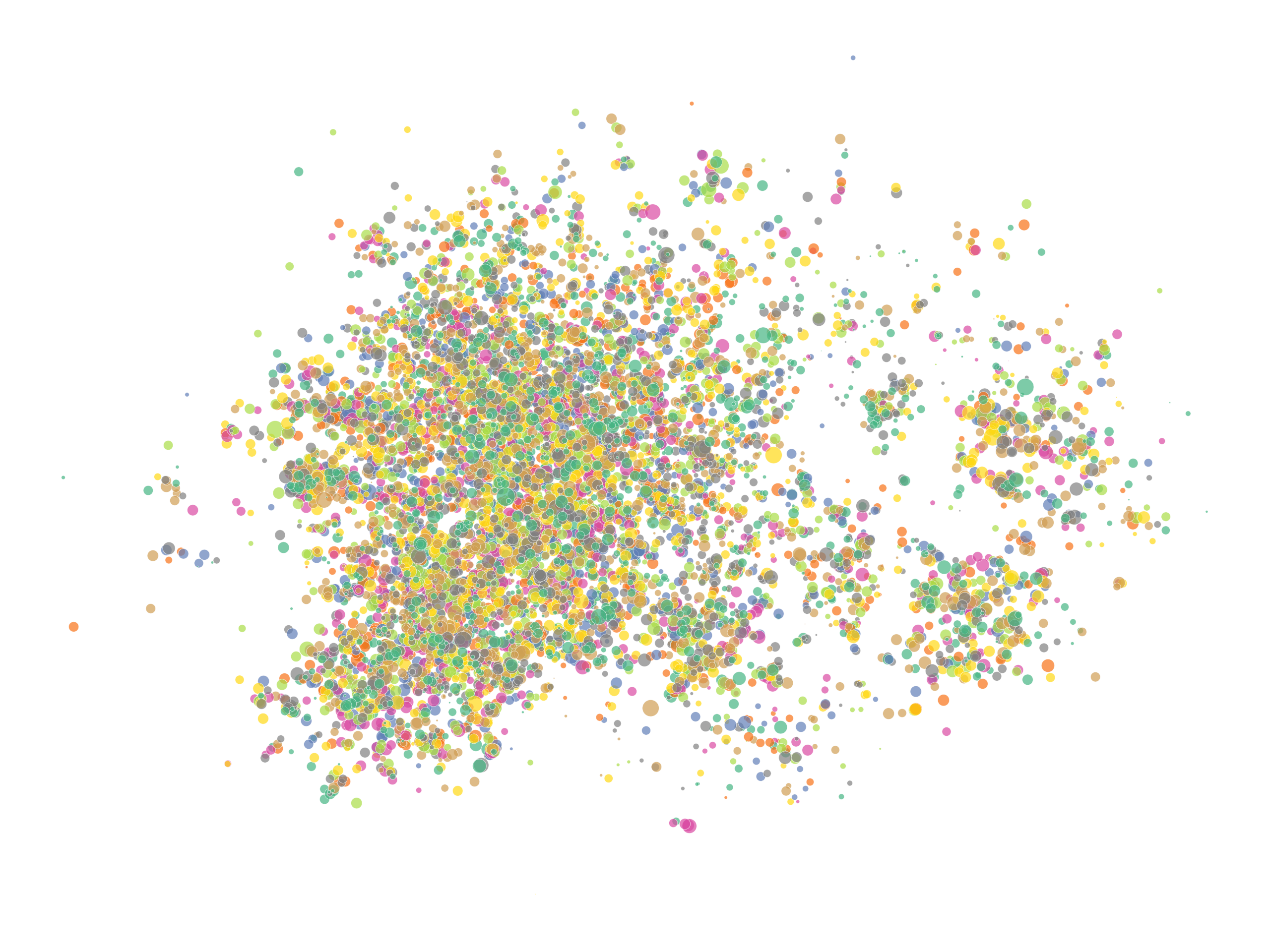}
    \caption{\textsl{SoSci} 2008}
    \label{fig:soc2008}
  \end{subfigure}
  \hfill
  \begin{subfigure}[t]{0.24\linewidth}
    \includegraphics[width=\linewidth]{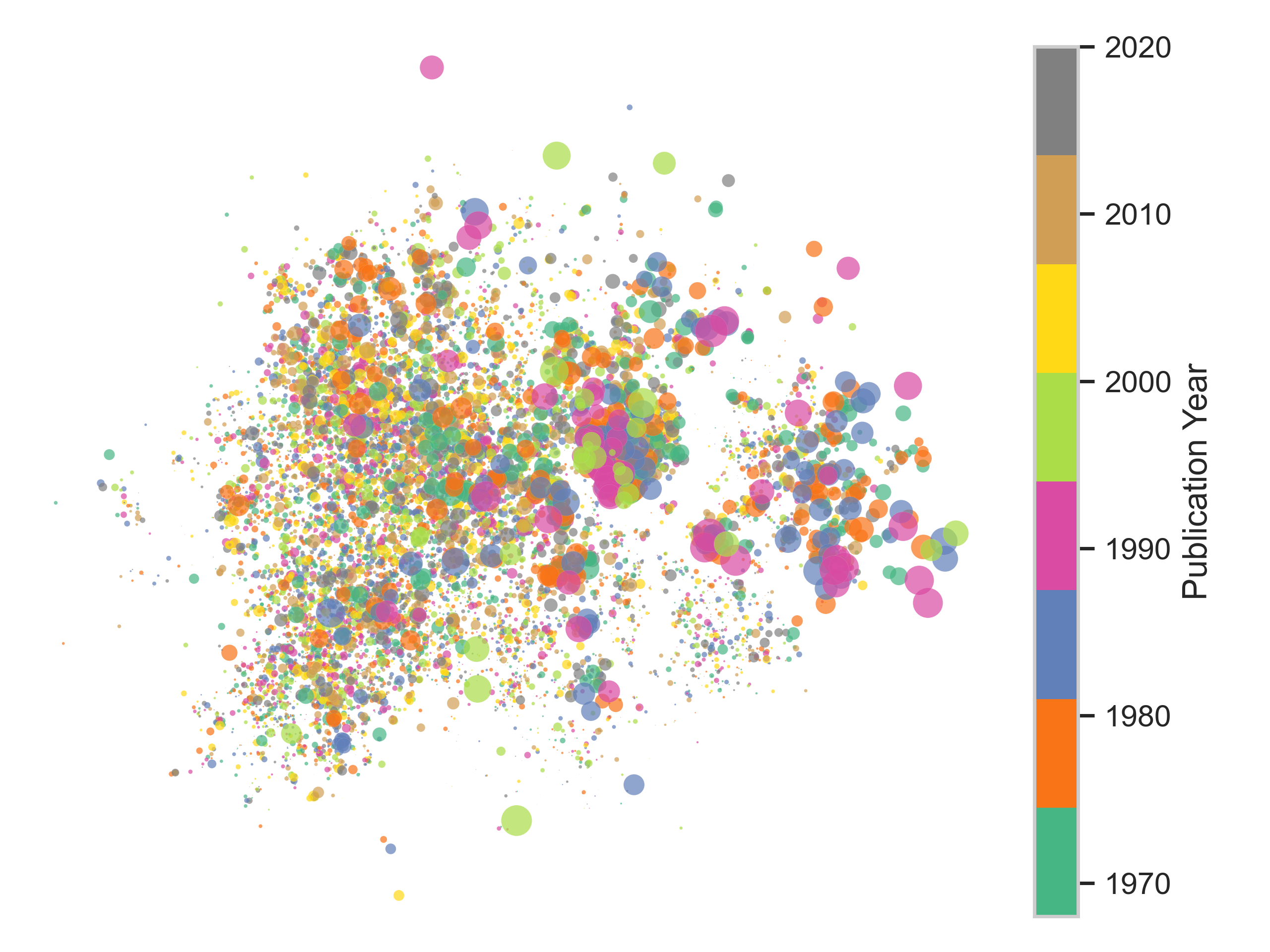}
    \caption{\textsl{SoSci} 2023}
    \label{fig:soc2023}
  \end{subfigure}
  \hfill
  \begin{subfigure}[t]{0.24\linewidth}
    \centering
    \includegraphics[width=\linewidth]{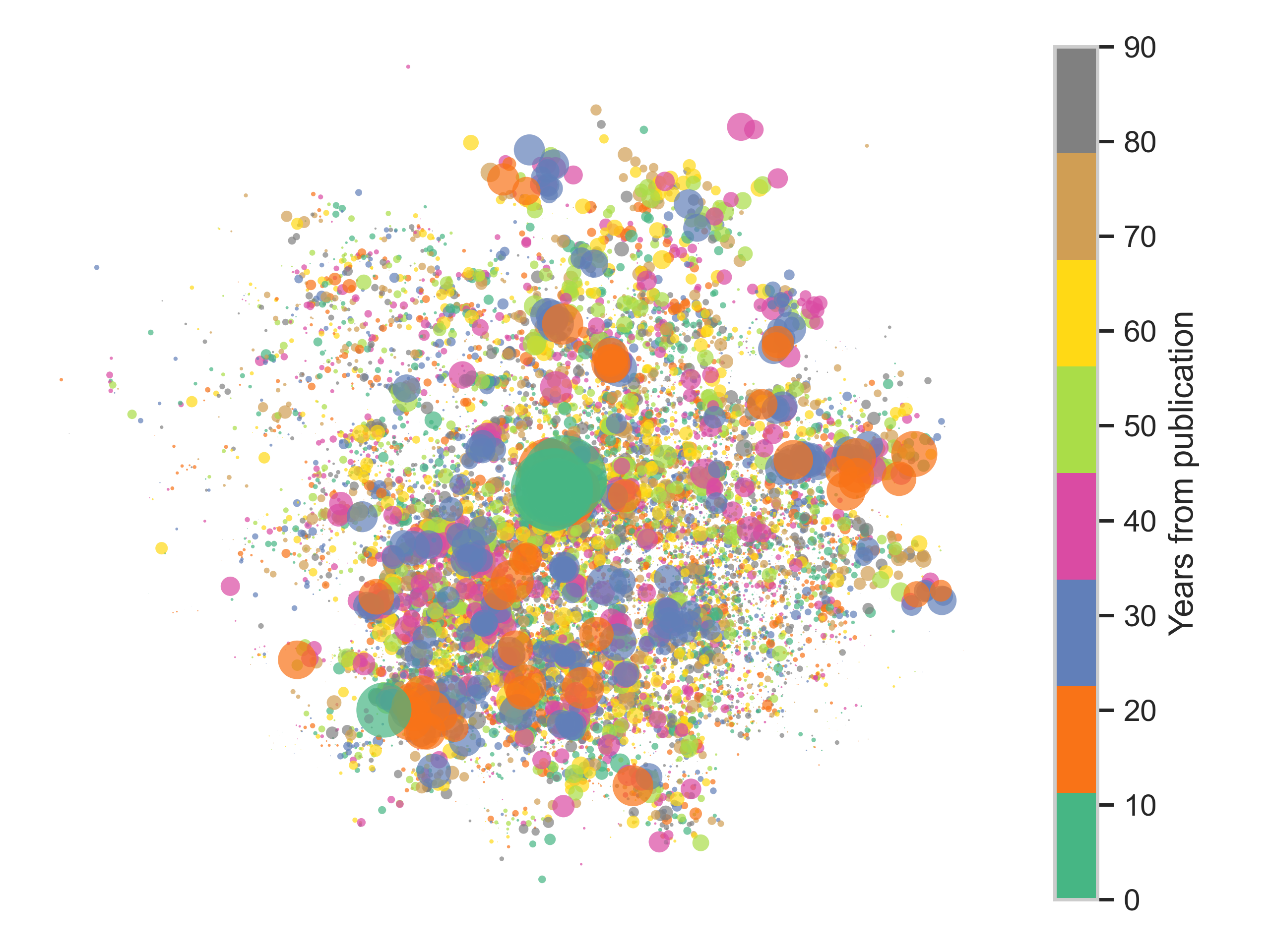}
    \caption{\textsl{Phys} 2023}
    \label{fig:phys}
  \end{subfigure}
  \caption{\textsc{\modelabbrev} 2-dimensional embedding space \textsc{Log-Normal}: \textsl{Social Science} (a)-(c) field evolution throughout the years with nodes color-coded based on their publication year. \textsl{Physics} present day status (d) with nodes color-coded based on years passed from their publication. Node sizes are based on each paper's mass, $f_i(t)\exp{(\alpha_i)}$.}
 \label{fig:history2}

\end{figure*}

\textbf{Impact quantification and space visualization}: We now continue by addressing the quality of paper impact characterization based on a target paper's incoming citation dynamics. In Figure \ref{fig:ML_ln}, we provide the inferred impact functions of the \textsc{\modelabbrev} and \textsc{FI-\modelabbrev}, under the \textsc{Log-Normal} and \textsc{Truncated} normal distributions. We further show the true impact dynamics through the citation histogram for each one of the corresponding papers. For the \textsc{Truncated} normal case we witness a general agreement between \textsc{\modelabbrev} and \textsc{FI-\modelabbrev} models, describing well the ground truth citation pattern. For the \textsc{Log-Normal} case, we witness a similar impact shape if the paper lifespan is short in terms of its incoming citation pattern. For larger lifespans \textsc{\modelabbrev} defines a much larger standard deviation than the \textsc{FI-\modelabbrev} returning heavier tails. In general, when compared to the true citation histogram both models using the \textsc{Log-Normal} distribution provide much heavier tails when the paper lifespan exceeds the $2$-year threshold. 
Nevertheless, the \textsc{Log-Normal} heavier tails may be more appropriate for future impact predictions (as papers stay "alive" longer).

Finally, we here provide embedding space visualizations of the target papers, accounting for their temporal impact in terms of their mass at a specific time point. Analytically, Figures \ref{fig:history} and \ref{fig:history2} (a)-(c) show the evolution of the embedding space for the domain of \textsl{Machine Learning} and \textsl{Social Sciences} from the year $1988$ until $2023$. We here observe how papers are published in a specific year and after they accumulate a specific amount of impact/mass they perish in the next years/snapshots of the network. Lastly, Figure \ref{fig:history2} (d) shows the present day status of the \textsl{Physics} based on the years passed from the publication for each paper. 

\textbf{Broader scope:} Our proposed approach has a general aim of providing a principled analysis of single-event dynamic networks, reconciling impact characterization with graph representation learning. We highlight here that the impact function can further be extended to kernel density estimation approaches, non-parametric methods, and mixture models. As an example, in Figure \ref{fig:mix_main} we provide an example of a mixture-model for the \textsl{Art} network. We witness how more advanced methods can accurately describe more complex citation patterns. This showcases the generalization power of \textsc{DISEE} which does not solely rely on a specific choice of impact function. 

\section{Conclusion}\label{sec:conclusion}
We have proposed the \textsc{D}ynamic \textsc{I}mpact \textsc{S}ingle-\textsc{E}vent \textsc{E}mbedding Model (\textsc{\modelabbrev}), a reconciliation between traditional impact quantification approaches with a Latent Distance Model (\textsc{LDM}). We have focused on Single-Event Networks (\textsc{SEN}s), and more specifically in citation networks, where we to the best of our knowledge for the first time derived and explored the Single-Event Poisson Process (SE-PP). Such a process defines an appropriate likelihood allowing for a principled analysis of \textsc{SEN}s. In order to define powerful ultra-low dimensional network embeddings, we turned to the representation power of the \textsc{LDM}. 
We introduced impact functions parameterized through an appropriate probability density function, such as the log-normal distribution. Through extensive experiments, we showed that the \textsc{\modelabbrev} inherited the link prediction performance of the powerful \textsc{LDM}. Furthermore, we showed how the temporal impact successfully characterized citation patterns, showcasing that the \textsc{\modelabbrev} model successfully reconciles powerful embedding approaches with citation dynamics impact characterization. Finally, visualizations of the embedding space clearly depict the lifecycle of target papers, highlighting their emergence, impact duration, and eventual decline as science evolves over the years.

An important aspect of SciSci is the future impact prediction for newly published/unseen (during training) papers. 
Predicting such citation patterns requires an inductive setting that can be obtained by extending our proposed \textsc{\modelabbrev} model defining paper embeddings via Graph Neural Networks based on the citing pattern (and potentially the inclusion of node features). We leave such formulations as a future research direction.

\clearpage
\section*{Acknowledgements}
We would like to express sincere appreciation and thank the reviewers for their constructive feedback and their insightful comments. We thank the Independent Research Fund Denmark for supporting this work [grant number: 0136-00315B].
\bibliography{reference.bib}

\end{document}